\documentclass[aps,prb,reprint,groupedaddress,showpacs]{revtex4-1}
\usepackage {graphicx}
\usepackage{amsmath}

\begin{document}

\title{Phase Transition of the Ising Model on Fractal Lattice}

\author{Jozef \textsc{Genzor}$^{1}$}
\author{Andrej \textsc{Gendiar}$^{1}$}
\email[]{andrej.gendiar@savba.sk}
\author{Tomotoshi \textsc{Nishino}$^{2}$}
\affiliation{$^1$Institute of Physics, Slovak Academy of Sciences, D\'ubravsk\'a cesta 9, SK-845 11, Bratislava, Slovakia}
\affiliation{$^2$Department of Physics, Graduate School of Science, Kobe University, Kobe 657-8501, Japan}

\date{\today}

\begin{abstract}
Phase transition of the Ising model is investigated on a planar lattice that has a fractal 
structure. On the lattice, the number of bonds that cross the border of a finite area is 
doubled when the linear size of the area is extended by a factor of four. The free energy 
and the spontaneous magnetization of the system are obtained by means of the higher-order
tensor renormalization group method. The system exhibits the order-disorder phase transition,
where the critical indices are different from that of the square-lattice Ising model. An
exponential decay is observed in the density matrix spectrum even at the critical point.
It is possible to interpret that the system is less entangled because of the fractal geometry.
\end{abstract}

\pacs{75.10.Pq, 75.10.Jm, 75.40.Mg}
\maketitle

%%%%
\section{Introduction}
%%%%

Phase transition and critical phenomena have been one of the central issues in statistical 
analyses of condensed matter physics~\cite{Domb_Green}. When the second-order phase 
transition is observed, thermodynamic functions, such as the free energy, the internal energy, 
and the magnetization, show non-trivial behavior around the transition temperature 
$T_{\rm c}^{~}$~\cite{Fisher, Stanley}. This critical singularity reflects the absence of any 
scale length at $T_{\rm c}^{~}$, and the power-law behavior of thermodynamic functions 
around the transition is explained by the concept of the renormalization 
group~\cite{Kadanoff, Kadanoff2, Wilson-Kogut,Domb_Green}.

Analytic investigation of the renormalization group flow in $\varphi^4_{~}$-model shows 
that the Ising model exhibits a phase transition when the lattice dimension is larger than 
one, which is the lower critical dimension~\cite{Wilson-Kogut, Justin}. In a certain sense, 
the one-dimensional Ising model shows rescaled critical phenomena around 
$T_{\rm c}^{~} = 0$. When the lattice dimension is larger than four, which is the upper 
critical dimension, and provided that the system is uniform, then the Ising model on regular 
lattices exhibits mean-field-like critical behavior.

Compared with critical phenomena on regular lattices, much less is known on fractal 
lattices. Renormalization flow is investigated by Gefen et al.,~\cite{Gefen1, Gefen2,
Gefen3, Gefen4} 
where correspondence between lattice structure and the value of critical indices is
not fully understood in a quantitative manner. For example, the Ising model on the Sierpinski 
gasket does not exhibit phase transition at any finite temperature, although the Hausdorff 
dimension of the lattice, $d_{\rm H}^{~} = \ln 3 / \ln 2 \approx 1.585$, is larger than one~\cite{Gefen5, Luscombe}.
The absence of the phase transition could be explained by the fact that the number of 
interfaces, i.e. the outgoing bonds from a finite area, does not increase when the size 
of the area is doubled on the gasket. A non-trivial feature of this system is that there is
a logarithmic scaling behavior in the internal energy toward zero temperature~\cite{log}. 
The effect of anisotropy has been considered recently~\cite{Ran}.
In case of the Ising model on the Sierpinski 
carpet, presence of the phase transition is proved~\cite{Vezzani}, and its critical indices 
were roughly estimated by Monte Carlo simulations~\cite{Carmona}. It should be noted 
that it is not easy to collect sufficient number of data plots for finite-size
scaling~\cite{FSS} 
on such fractal lattices by means of Monte Carlo simulations, because of the exponential 
blow-up of the number of sites in a unit of fractal. 

%%%%
\begin{figure}[tb]
\includegraphics[width=0.4\textwidth,clip]{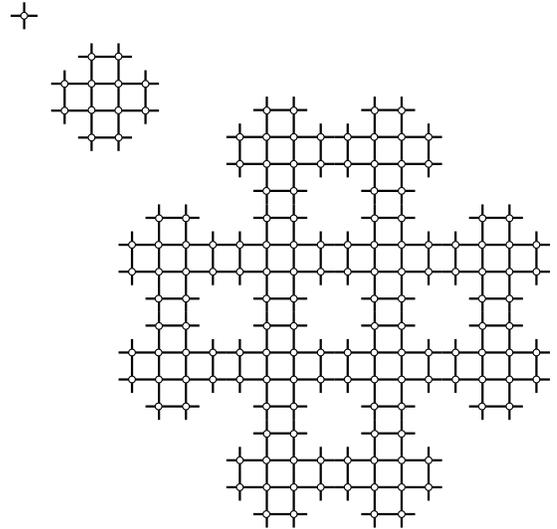}
\caption{
Composition of the fractal lattice. Upper left: a local vertex around an Ising spin
shown by the empty dot. Middle: the basic cluster which contains $N_1^{~} = 12$ 
vertices. Lower right: the extended cluster which contains $N_2^{~} = 12^2_{~}$ 
vertices. In each step of the system extension, the linear size of the system increases 
by the factor of 4, where only 12 units are linked, and where 4 units at the corners 
are missing, if it is compared with a 4 by 4 square cluster. 
}
\label{fig:Fig_1}
\end{figure}
%%%%

In this article, we investigate the Ising model on a planar fractal lattice, shown in 
Fig.~\ref{fig:Fig_1}. The lattice consists of vertices around the lattice points, which are 
denoted by the empty dots in the figure, where there are Ising spins. The whole lattice 
is constructed by recursive extension processes, where the linear size of the system 
increases by the factor of four in each step. If the lattice was a regular square one, 
$4 \times 4 = 16$ units are connected in the extension process, whereas only 12 units 
are connected on this fractal lattice; 4 units are missing in the corners. As a result, 
the number of sites contained in a cluster after $n$ extensions is $N_n^{~} = 12^n_{~}$, 
and the Hausdorff dimension of this lattice is $d_{\rm H}^{~} = \ln 12 / \ln 4 \approx 1.792$. 
The number of outgoing bonds from a cluster is only doubled in each extension process 
since the sites and the bonds at each corner are missing. If we evaluate the lattice dimension 
from the relation 
\begin{equation}
M = L^{d-1}_{~}
\end{equation}
between the linear dimension $L$ and the number of outgoing bonds $M$, we have $d = 1.5$, 
since $M$ is proportional to $\sqrt{L}$ on the fractal. Remark that the value is different from 
$d_{\rm H}^{~} \approx 1.792$

We report the critical behavior of the Ising model on the fractal lattice when the system 
size is large enough. Thermodynamic properties of the system are numerically studied by 
means of the Higher-Order Tensor Renormalization Group (HOTRG) method~\cite{HOTRG}.
The system exhibits the order-disorder phase transition, where the critical indices are 
different from the square lattice Ising model. In the next Section we introduce a 
representation of the Ising model in terms of a vertex model, which is suitable for
numerical analyses by means of the HOTRG method. In Sec.~III, we show the calculated result 
around the transition temperature $T_{\rm c}^{~}$. Conclusions are summarized in the last 
Section.

\section{Vertex representation}

We introduce a representation of the Ising model as a (symmetric) 16-vertex model. 
The Ising interaction between two adjacent Ising spins $\sigma$ and $\sigma'$,
where each one takes either $+1$ or $-1$, is represented by the diagonal Hamiltonian
\begin{equation}
H( \sigma, \sigma' ) = - J \sigma \sigma' \, ,
\end{equation}
where $J > 0$ represents the ferromagnetic coupling. Throughout this article we 
assume that there is no external magnetic field. The corresponding local Boltzmann 
weight on the bond is given by
\begin{equation}
\exp\left[ - \frac{ H( \sigma, \sigma' ) }{ k_{\rm B} T } \right] = 
\exp\left[ \frac{ J }{ k_{\rm B} T } \, \sigma \sigma' \right] = e^{K \sigma \sigma'}_{~} \, ,
\end{equation}
where $T$ is the temperature, $k_{\rm B}$ is the Boltzmann constant, and we have
introduced a parameter $K = J / k_{\rm B} T$.

It is possible to factorize the bond weight $e^{K \sigma \sigma'}_{~}$ into two parts, 
by introducing an auxiliary spin $s = \pm 1$, which is often called an `ancilla', and 
which is located between $\sigma$ and $\sigma'$~\cite{Fisher_M}.
A key relation is
\begin{equation}
e^{K \sigma \sigma'}_{~} =
\frac{1}{2 \left( \cosh 2 {\overline K} \right)^{1/2}_{~} } \, 
\sum_s^{~} \, e^{{\overline K} s ( \sigma + \sigma' )}_{~} \, ,
\label{expKss}
\end{equation}
where the r.h.s. takes the value 
$\left( \cosh 2 {\overline K} \right)^{1/2}_{~}$ when $\sigma = \sigma'$, and 
$\left( \cosh 2 {\overline K} \right)^{-1/2}_{~}$ when $\sigma \neq \sigma'$,
and where Eq.~\eqref{expKss} holds under the condition
\begin{equation}
e^{K}_{~} = \left( \cosh 2 {\overline K} \right)^{1/2}_{~} \, .
\end{equation}
The new parameter ${\overline K}$ is then expressed as follows
\begin{equation}
e^{\overline K}_{~} = \sqrt{ e^{2K}_{~} + \sqrt{ e^{4K}_{~} - 1 } } \, .
\end{equation}
Thus, if we introduce a factor
\begin{equation}
W_{\sigma s}^{~} = e^{\overline K \sigma s}_{~} \left[ 2 \left( \cosh 2 {\overline K} \right)^{1/2}_{~} \right]^{-1/2}_{~}
\label{factW}
\end{equation}
for each division of a bond, we can rewrite the Ising interaction in the following form
\begin{equation}
e^{K \sigma \sigma'}_{~} 
= \sum_{s}^{~} \, W_{\sigma s}^{~} \, W_{\sigma' s}^{~} \, .
\label{eKWW}
\end{equation}

By means of the factorization in Eq.~\eqref{eKWW}, we can map the square-lattice 
Ising model into the symmetric 16-vertex model, where the local vertex weight is 
defined as
\begin{equation}
T_{s\,s's''s'''}^{~} = \sum_{\sigma}^{~} \, W_{\sigma s}^{~} \, W_{\sigma s'}^{~} \, W_{\sigma s''}^{~} \, W_{\sigma s'''}^{~} \, .
\label{TWWWW}
\end{equation}
In the upper-left corner of Fig.~\ref{fig:Fig_1}, we have shown the graphical representation of
the vertex weight $T_{s\,s's''s'''}^{~}$, where the open circle denotes the Ising spin $\sigma$, 
which is summed up. The four short bars around the Ising spin in Fig.~\ref{fig:Fig_1} show the halves of 
the bonds, where there are auxiliary spins $s$, $s'$, $s''$, and $s'''$ at the end of each short bar. 

In case we consider a finite-size cluster with rectangular shape with free boundary conditions, 
we have to prepare a new boundary Boltzmann weight
\begin{equation}
P_{s\,s's''}^{~} = \sum_{\sigma}^{~} \, W_{\sigma s}^{~} \, W_{\sigma s'}^{~} \, W_{\sigma s''}^{~} 
\label{Psss}
\end{equation}
and a corner Boltzmann weight
\begin{equation}
C_{s\,s'}^{~} = \sum_{\sigma}^{~} \, W_{\sigma s}^{~} \, W_{\sigma s'}^{~}  \, .
\label{Css}
\end{equation}
It should be noted that these boundary weights $P_{s\,s's''}^{~}$ and $C_{s\,s'}^{~}$ are 
obtained by taking partial trace for the vertex weight; we have the  relations
\begin{equation}
P_{s\,s's''}^{~} = 
\frac{\sum_{s'''}^{~} \, T_{s\,s's''s'''}^{~}}{\sum_{s'''}^{~} \, W_{\sigma s'''}^{~}}
\end{equation}
and
\begin{equation}
C_{s\,s'}^{~} = 
\frac{\sum_{s''s'''}^{~} T_{s\,s's''s'''}^{~}}{
( \sum_{s''}^{~} \, W_{\sigma s''}^{~} ) 
( \sum_{s'''}^{~} \, W_{\sigma s'''}^{~} )} \, ,
\end{equation}
where one can neglect the denominator when one is interested in the critical singularity;
the denominators just subtract a regular function from the free energy of the system.
In case that one needs fixed boundary conditions, it is sufficient to avoid taking the
configuration sum for $\sigma$ in the r.h.s. of both Eq.~\eqref{Psss} and Eq.~\eqref{Css}, 
and to set all the boundary spins to be either $+1$ or $-1$ according to the condition. 
The vertex weights $T_{s\,s's''s'''}^{~}$, $P_{s\,s's''}^{~}$, and $C_{s\,s'}^{~}$ are invariant 
under arbitrary permutation of the indices.

There are various choices of the factorization of the bond weight in Eq.~\eqref{eKWW}. 
Instead of using the relation in Eq.~\eqref{factW}, one can introduce an asymmetric 
decomposition
\begin{equation}
W =  \left(\begin{array}{cc} 
\sqrt{\cosh K}, & \sqrt{\sinh K}   \\
\sqrt{\cosh K} ,& -\sqrt{\sinh K} \end{array} \right) \, ,
\label{Tasym}
\end{equation}
where we have used the matrix notation for the weight $W_{\sigma s}^{~}$. This
expression is often employed in the tensor network formulations~\cite{HOTRG},
which does not require any typical symmetry for local weights, as long as the numerical
treatment is concerned. In case this asymmetric factorization in Eq.~\eqref{Tasym} is 
employed, one has to care about the order of the indices in $T_{s\,s's''s'''}^{~}$~\cite{RG}.
In the following numerical calculation, we use the symmetric factorization.

The fractal lattice we treat in this article is constructed by a recursive joining
process of the local tensors, which is nothing but a vertex weight in 
Eq.~\eqref{TWWWW} at the beginning. In each extension process, we join 12 local 
tensors as shown in the middle of Fig.~\ref{fig:Fig_1}. In the joining process, we take 
the configuration sum for those tensor indices inside the cluster, leaving those on the 
border that become new tensor indices of the extended tensor. Because of the fractal 
geometry, some of the bonds inside the cluster are not connected with each other. We 
also take configuration sum for these dangling bonds, and the process just change the 
normalization of the partition function by amount of 
\begin{equation}
\sum_s^{~} \, W_{\sigma s}^{~} = 
2 \cosh {\overline K} \, 
\left[ 2 \left( \cosh 2 {\overline K} \right)^{1/2}_{~} \right]^{-1/2}_{~}
\end{equation}
for each, if we choose the definition of $W_{\sigma s}^{~}$ in Eq.~\eqref{factW}. 
We take the rescaling effect into account, although the rescaling is not essential to the 
thermodynamic properties of the system, in particular to its critical singularity.
In this manner, what we are dealing with is the Ising model, where there are only 
spins denoted by the empty dots in Fig.~\ref{fig:Fig_1}.

At first we have only 4 spins $s$, $s'$, $s''$, and $s'''$ on the outgoing bonds, and
after $n$ extensions of the system, we have $4 \times 2^n_{~}$ border spins on the 
{\it surface} of the extended cluster. The application of the HOTRG to this fractal 
system is straightforward. The recursive structure of the  lattice is suitable for the 
repeated process of system extensions and renormalization group transformations 
in the HOTRG method. The partition function $Z_n^{~}( T )$ of the system after $n$ 
extensions is obtained by the contraction of the extended tensors; we choose the 
periodic boundary conditions to evaluate 
\begin{equation}
Z_n^{~}( T ) = \sum_{ij}^{~} \, T_{ijij}^{(n)} \, ,
\end{equation}
where $T_{ijkl}^{(n)}$ is the renormalized local tensor obtained after $n$ extensions.

\section{Numerical Results}

In order to simplify the numerical analysis, we choose the parameterization $J = k_{\rm B} = 1$, 
and thus we have $K = 1 / T$. In the numerical calculation by means of HOTRG, 
we keep $D = 24$ states at most for block spin variables. We have verified that 
the choice $D = 24$ is sufficient for obtaining the converged free energy 
\begin{equation}
F_n^{~}( T ) = - k_{\rm B} T \ln Z_n^{~}( T )
\end{equation}
in the entire temperature region~\cite{chi}. 
We treat the free energy per site
\begin{equation}
f( T ) = \lim_{n \rightarrow \infty}^{~} \, \frac{F_n^{~}( T )}{N_n^{~}}
\end{equation}
in the following thermodynamic analyses, where the r.h.s. converges 
already for $n \lesssim 30$ .

\begin{figure}[tb]
\includegraphics[width=0.48\textwidth,clip]{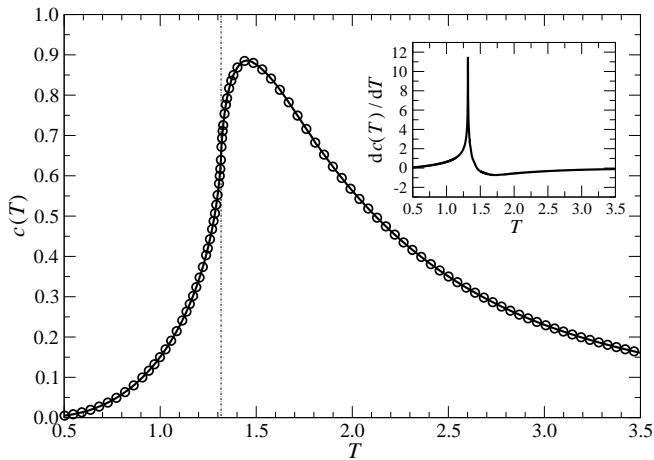}
\caption{The specific heat $c( T )$ per site in Eq.~\eqref{Cv}. 
Inset: numerical derivative of $c( T )$ with respect to temperature;
a sharp peak is observed at $T_{\rm c}^{~} \approx 1.317$.}
\label{fig:Fig_2}
\end{figure}

Figure~\ref{fig:Fig_2} shows the temperature dependence of the specific heat per site 
\begin{equation}
c( T ) = \frac{\partial}{\partial T} \, u( T ) \, ,
\label{Cv}
\end{equation}
where $u( T )$ is the internal energy per site
\begin{equation}
u( T ) = - T^2_{~} \, \frac{\partial}{\partial T} \, \frac{f( T )}{T} \, ,
\end{equation}
and the temperature derivatives are performed numerically. There is no 
singularity in $c( T )$ around its maximum. One might find a weak non-analytic 
behavior at $T_{\rm c}^{~} \approx 1.317$, which is marked by the dotted line in the figure; 
the numerical derivative of $c( T )$ with respect to temperature (plotted in the inset)
has a sharp peak at the critical temperature $T_{\rm c}^{~}$. 
It is, however, difficult to determine the critical exponent $\alpha$ precisely, because of the
weakness in the singularity; as shown in the figure, $c( T )$ around $T_{\rm c}^{~}$ is almost 
linear in $T$, and therefore $\alpha$ is nearly zero.

%%%%
%%%%

\begin{figure}[tb]
\includegraphics[width=0.48\textwidth,clip]{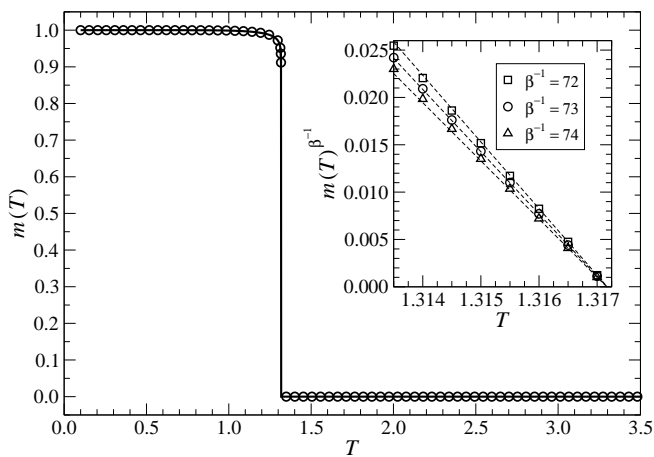}
\caption{The spontaneous magnetization per site $m( T )$. 
Inset: the power-law behavior below $T_{\rm c}^{~}=1.31716$.}
\label{fig:Fig_3}
\end{figure}

Figure~\ref{fig:Fig_3} shows the spontaneous magnetization per site $m( T )$, which is obtained
by inserting a $\sigma$-dependent local weight 
\begin{equation}
{\tilde T}_{s \, s' s'' s'''}^{~} = \sum_{\sigma}^{~} \, 
\sigma \, W_{\sigma s}^{~} \, W_{\sigma s'}^{~} \, W_{\sigma s''}^{~} \, W_{\sigma s'''}^{~} 
\end{equation}
into the system. Since the fractal lattice is inhomogeneous, the value is weakly 
dependent on the location of the observation site, but the critical behavior is not
affected by the location; we choose a site from the four sites that are in the middle 
of the 12-site cluster shown in Fig.~\ref{fig:Fig_1}. The numerical calculation by HOTRG 
captures the spontaneous magnetization $m( T )$ below $T_{\rm c}^{~}$ since any tiny 
round-off error is sufficient for breaking the symmetry inside low-temperature ordered 
state.  Around the transition temperature, the magnetization satisfies a power-law behavior
\begin{equation}
m( T ) \propto
% | T_{\rm c}^{~} - T |^{1/73}_{~} \, ,
| T_{\rm c}^{~} - T |^{0.0137}_{~} \, ,
\end{equation}
where the precision of the exponent is around 2\%, 
which can be read out from the inset of Fig.~\ref{fig:Fig_3}
as a tiny deviation from the linear dependence (the dashed lines) in $m(T)^{1/\beta}$ near
$T_{\rm c}^{~}$.

%%%%
%%%%

%%%%
\begin{figure}[tb]
\includegraphics[width=0.48\textwidth,clip]{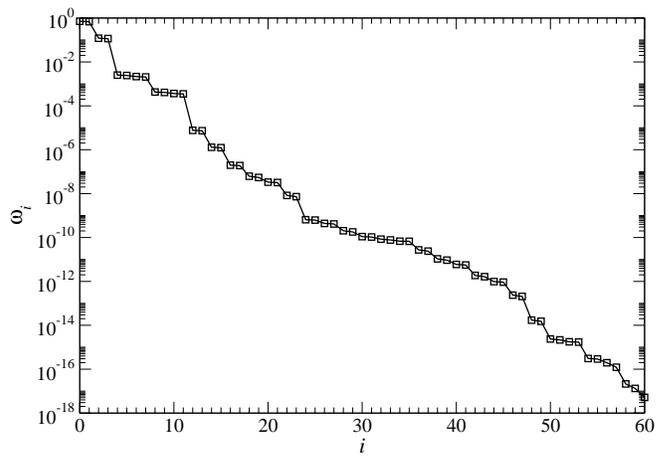}
\caption{Decay of the singular values after $n=8$ extensions.}
\label{fig:Fig_4}
\end{figure}
%%%%

As a byproduct of the numerical HOTRG calculation, we can roughly
observe the entanglement spectrum,~\cite{entanglement} which is the distribution of
the eigenvalue $\omega_i^{~}$ of the density matrix that is created for the purpose of 
obtaining the block spin transformation. Since the effect of  {\it environment} is not considered in our 
implementation of the HOTRG method, the eigenvalue $\omega_i^{~} = \lambda^2_i$ is 
obtained as the square of the singular values $\lambda_i^{~}$ in the higher-order
singular value decomposition applied to the extended tensors.
Figure~\ref{fig:Fig_4} shows $\omega_i^{~}$ at 
$T = T_{\rm c}^{~}$ in the decreasing order. The  decay is rapid, and therefore further 
increase of the number of block-spin state from $D = 24$ to a larger number does not 
significantly improve the precision in $Z_n^{~}$; the difference in $f( T_{\rm c}^{~} )$
between $D = 8$ and $D = 16$ is already of the order of $10^{-6}_{~}$.
It should be noted that the eigenvalues are not distributed equidistantly in logarithmic
scale; the corner double line structure is absent~\cite{CDL1, CDL2}.

\section{Conclusions and Discussions}

We have investigated the Ising model on the fractal lattice shown in Fig.~\ref{fig:Fig_1}
by means of the HOTRG method. The calculated specific heat $c( T )$ suggests that the model
shows 2$^{\rm nd}$ order phase transition. Qualitatively speaking, the presence of weak singularity
in the specific heat agrees with the result of the $\epsilon$-expansion, which shows the
increasing nature of the critical exponent in $c(T)$ with respect to the space dimension
$d~$~\cite{Wilson-Kogut}. The calculated spontaneous magnetization $m( T )$ also supports
the 2$^{\rm nd}$ order phase transition with the exponent $\beta_{\rm fractal}^{~}
\approx 0.0137$, which is smaller by one order of magnitude than the critical exponent
$\beta_{\rm square}^{~} = 1/8 = 0.125$ of the square-lattice Ising model.

The fractal structure of the lattice modifies the entanglement spectrum from 
that on the square lattice explained by the corner double line 
picture~\cite{CDL1, CDL2}. Since each corner is missing in the fractal 
structure in Fig.~\ref{fig:Fig_1}, short-range entanglement is almost filtered out in the
process of the renormalization group transformation. This may be the reason 
why we do not need many degrees of freedom for the renormalized tensors.
The situation is similar to the entanglement structure 
reported in the tensor network renormalization~\cite{TNR1, TNR2, TNR3, TNR4, TNR5, Loop}.

The lattice geometry of the fractal lattice can be modified in several manners.
For example, one can alternate the system extension process of the fractal for
the purpose of modifying the Hausdorff dimension; for every odd $n$ the extension 
with 12 vertices shown in Fig.~\ref{fig:Fig_1} is performed, and for even $n$ the 
normal extension with 16 vertices on the square-lattice is performed. Alternatively,
one  can also modify the basic cluster, in such a manner as introducing 6 by 6 cluster 
where 4 corners are missing, etc. It is also worth considering three-dimensional 
fractal lattice, and apply the HOTRG method as it was done for the cubic lattice
Ising model~\cite{Xie}. These modifications do not spoil the applicability of the HOTRG 
method while the numerical requirement is heavier than the current research. 
Analyses of quantum systems on a variety of fractal lattice is another possible
extensions~\cite{Voigt1, Voigt2}. These further study may clarify the role of the entanglement 
in the universality of the phase transition in both regular and fractal lattices.

\begin{acknowledgments}
This work was supported by the projects VEGA-2/0130/15 and QIMABOS APVV-0808-12.
T.~N. and A.~G. acknowledge the support of Grant-in-Aid for Scientific Research.
\end{acknowledgments}


\begin{thebibliography}{99}
\bibitem{RG} The symmetry in the local tensors is not always preserved when one performs
the renormalization group transformation in the HOTRG method. Thus, for most of the cases, the 
symmetry is not that important in the numerical calculations.
\bibitem{chi} A larger value of $D$ is necessary if one needs tiny density-matrix eigenvalues for the 
purpose of analyzing their asymptotic decay.
\bibitem{entanglement} It is possible to identify the system boundary of a finite area of 2d classical 
lattice models as `a wave function' of a certain 1d quantum system. In this manner one naturally 
finds the quantum-classical correspondence, and can introduce the notion of entanglement in classical 
lattice models.
\end{thebibliography}
\end{document}